\documentclass[11pt, a4paper, logo, onecolumn, nonumbering, address]{deepmind}
\usepackage{kantlipsum, lipsum}
\usepackage{booktabs}       

\usepackage{siunitx}
\usepackage{array}
\usepackage{boldline}
\usepackage{xcolor}
\usepackage{wrapfig}
\usepackage{enumitem}
\usepackage{subfigure}
\usepackage{wrapfig}
\usepackage{blindtext}
\newcolumntype{?}{!{\vrule width 2pt}}

\usepackage{multirow}

\usepackage{caption} 
\usepackage[authoryear, sort&compress, round]{natbib} 
\usepackage{mathbbol}
\newcommand{\R}{\pmb{\mathbb{R}}}

\usepackage{tikz}

\usetikzlibrary{shapes,decorations,arrows,calc,arrows.meta,fit,positioning}
\tikzset{
    -Latex,auto,node distance =1 cm and 1 cm,semithick,
    state/.style ={ellipse, draw, minimum width = 0.7 cm},
    point/.style = {circle, draw, inner sep=0.04cm,fill,node contents={}},
    bidirected/.style={Latex-Latex,dashed},
    el/.style = {inner sep=2pt, align=left, sloped}
}

\usepackage{url}            
\usepackage{booktabs}       
\usepackage{amsfonts}       
\usepackage{nicefrac}       
\usepackage{microtype}      

\usepackage{svg}
\usepackage{gensymb}
\usepackage{enumitem}

\begin{document}

\title{Model-free conventions in multi-agent reinforcement learning with heterogeneous preferences}  

\reportnumber{} 

\renewcommand{\today}{October 2020}

\author[1, 2]{Raphael K{\"o}ster}
\author[1]{Kevin R. McKee}
\author[1]{Richard Everett}
\author[1]{Laura Weidinger}
\author[1]{William S. Isaac}
\author[1]{Edward Hughes}
\author[1]{Edgar A. Du\'e\~nez-Guzm\'an}
\author[1]{Thore Graepel}
\author[1]{Matthew Botvinick}
\author[1, 2]{Joel Z. Leibo}

\affil[1]{DeepMind}
\affil[2]{Corresponding authors: rkoster@google.com; jzl@google.com}

\correspondingauthor{rkoster@google.com; jzl@google.com}

\begin{abstract}
Game theoretic views of convention generally rest on notions of common knowledge and hyper-rational models of individual behavior. However, decades of work in behavioral economics have questioned the validity of both foundations. Meanwhile, computational neuroscience has contributed a modernized “dual process” account of decision-making where model-free (MF) reinforcement learning trades off with model-based (MB) reinforcement learning. The former captures habitual and procedural learning while the latter captures choices taken via explicit planning and deduction. Some conventions (e.g. international treaties) are likely supported by cognition that resonates with the game theoretic and MB accounts. However, convention formation may also occur via MF mechanisms like habit learning; though this possibility has been understudied. Here, we demonstrate that complex, large-scale conventions can emerge from MF learning mechanisms. This suggests that some conventions may be supported by habit-like cognition rather than explicit reasoning. We apply MF multi-agent reinforcement learning to a temporo-spatially extended game with incomplete information. In this game, large parts of the state space are reachable only by collective action. However, heterogeneity of tastes makes such coordinated action difficult: multiple equilibria are desirable for all players, but subgroups prefer a particular equilibrium over all others. This creates a coordination problem that can be solved by establishing a convention. We investigate start-up and free rider subproblems as well as the effects of group size, intensity of intrinsic preference, and salience on the emergence dynamics of coordination conventions. Results of our simulations show agents establish and switch between conventions, even working against their own preferred outcome when doing so is necessary for effective coordination.

\end{abstract}

\maketitle
\balance


\emph{``Two men, who pull the oars of a boat, do it by an agreement or convention, tho' they have never given promises to each other. Nor is the rule concerning the stability of possessions the less derived from human convention, that it arises gradually, and acquires force by a slow progression, and by our repeated experience of the inconvenience of transgressing it. ... In like manner are languages gradually establish'd by human conventions without any promise. In like manner do gold and silver become the common measures of exchange, and are esteem'd sufficient payment for what is of a hundred times their value.''}

\begin{flushright}
David Hume, \emph{A Treatise of Human Nature}~\citep{hume1739treatise}
\end{flushright}

\section{Introduction}
Conventions have causal force~\citep{bicchieri2006grammar, lewis1969convention}. For example, we dynamically conform in which side of the road to drive on, dependent on the way others drive. In Britain this means driving on the left, while in North America it means driving on the right. David Hume \citep{hume1739treatise}, and modern writers in his tradition \citep{binmore1994game, lewis1969convention, skyrms2014evolution, sugden1986economics, vanderschraaf2018strategic}, proposed that the broad organizing concepts of human society like property and justice are likewise conventional. But how can populations discover such useful conventions for organizing social life? Societies consist of individuals with diverse and often conflicting intrinsic preferences (tastes) over collective states. Even when subsets of individuals agree on a societal goal, they face a collective action problem~\citep{olson1965logic}. This is exacerbated by large state spaces and incomplete information about environment dynamics, as well as activities and preferences of other individuals. One implication of this setting, which arguably reflects the real world, is that discovering useful conventions entails a difficult problem of joint exploration.

Deep exploration---the kind that builds truly novel behaviors---is intractable in large worlds~\citep{kearns2002near, osband2019deep}. As anyone who ever tried to train a pigeon to bowl will attest, if you must wait to provide the first reward until the pigeon naturally emits an approximation of the desired behavior, you will have to wait a very long time indeed~\citep{peterson2004day}. In terms of Bayesian decision theory, the prior distribution must contain the novel behavior within its support. Otherwise no amount of evidence for its superiority will suffice to nudge its probability off zero~\citep{kalai1993rational}. Savage illustrated the problem with a pair of proverbs. A small world is one where you can always ``look before you leap''. A large world is one where you must sometimes ``cross that bridge when you come to it''~\citep{binmore2007rational, savage1951foundations}. Bayesian decision theory is only valid in small worlds. \emph{Prima facie}, multi-agent interaction would seem to exacerbate this issue since it inflates enormously the space of possible behaviors~\citep{allis1994searching}. Acquiring novel conventions to structure human behavior appears hopeless.

However, incentive structures are often such that simultaneous learning is likely to become attracted to a particular equilibrium out of a set of candidates. Conventions are such phenomena. How conventions are discovered follows from the learning algorithms of the agents. Most of the literature on convention formation takes one of two main approaches. On the one hand, conventions could emerge from rational calculation and strategic planning based on mutual knowledge of incentives like when multiple nations switched to use the Euro as their currency. This is the approach taken by classical game theory (e.g.~\cite{luce1957games}). On the other hand, convention formation could be treated as an emergent phenomenon driven by evolutionary forces, as in folk historical accounts of how currency came to replace barter in ancient societies. This is the approach taken by evolutionary game theory (e.g.~\cite{weibull1997evolutionary}). An objective of the present paper is to point out that a fraction\footnote{We don't know how large this fraction is, but there is no reason to believe it is small.} of real-world convention formation behavior is not well described by models based on either of these approaches. We show here the viability of an alternative approach to convention formation, grounded in theories of habit learning from neuroscience. 

A large body of work has examined the contrast between model-based (MB) and model-free (MF) reinforcement learning. MB reinforcement learning can prospectively plan to achieve desired outcomes using knowledge of the likely consequences of actions. In contrast, MF reinforcement learning gradually updates cached values of actions retrospectively from experience~\citep{dayan2008decision,  akam2015}. In particular, this computational dichotomy has been mapped onto many dual-process theories in psychology and neuroscience~\citep{dayan2009goal, collins2020beyond}. MF and MB reinforcement learning differ in their capacity for rapid revaluation. If an outcome that was once valued suddenly becomes devalued, then the MB decision-maker would immediately change their behavior while a MF decision-maker would have to relearn from additional repetitions of the experience~\citep{dickinson1985actions}. Similarly, a task may pit MB and MF based learning against each other by offering opportunities available only to learners who are sensitive to the transition structure of the task states ~\citep{DAW20111204}. In the brain, MF learning is associated with the dorsolateral striatum while MB learning is associated with prefrontal cortex~\citep{daw2005uncertainty}. In humans, culture establishes the habits and rituals that structure daily life (in part) via experience dependent plasticity in the striatum~\citep{graybiel2008habits}.

Habits occupy a middle ground between instinct and reason. In classical game theory, the conventions governing human society are often described in terms that resonate most with the MB account. Thomas Schelling provided a classic example in an experiment where he asked participants to imagine playing a game in which they had to meet one another in New York City without arranging a location in advance or communicating. Despite the enormous action space (they could pick any location in the city), most participants chose the same location: grand central station~\citep{schelling1960strategy}. This solution relies on mutual knowledge and prospective reasoning to succeed in one shot. Similarly, a computer Go program can employ self-play using its preprogrammed knowledge of the rules of the game---a perfect model---to progressively improve its skills until it can defeat the world champion~\citep{brown1951iterative, silver2017mastering, tesauro1994td}. MB control works prospectively, and in accord with the principles of Bayesian decision theory. This means mentally simulating the tree of possibilities and selecting the action believed to lead to the best long-term outcome\footnote{At least up to computation constraints. To go further, MB control relies on ``irrational'' heuristics~\citep{huys2015interplay}. In the particular case of Schelling's example, clearly some mutual knowledge of the salience of the particular solution, grand central station, was required. In this case, all the participants in Schelling's experiment were students at Yale university. So it's very likely that they would normally access New York City via the train from New Haven, which stops at Grand Central, thus making it the logical meeting place.}. In contrast, evolutionary forces operate by retrospective selection. They are underpinned either by instinct and natural selection (e.g.~\cite{gintis2007evolution, stake2004property}) or imitation and cultural selection (e.g.~\cite{sen2007emergence, young1996economics}). MF mechanisms are different. Like evolutionary selection, MF learning is retrospective. It operates on stored values which change gradually on the basis of new experience. However, unlike evolution, MF learning iteratively adjusts a system of interlocking predictions of future value towards self-consistency~\citep{sutton2018reinforcement}.

In what situations might a MF account of conventions be applicable? Consider the following: A jazz band decides to incorporate a new member. They practice and rehearse together to achieve coordination before performing their first gig. Ballroom dance partners practice a complex dance for months before a performance. A football team practices before the big game. Several different online social networks or messaging apps exist, but everyone comes to use the same social network or messaging app. And, David Hume's example: you and I can row a boat together. We need not even speak the same language. We could coordinate our rowing by trial and error~\citep{hume1739treatise}. All these examples concern the formation of conventions. Notice however that these conventions differ from Schelling's example in that they are insensitive to rapid revaluations. If the wind conditions change in Hume's rowboat both rowers must trial-and-error their way to a new equilibrium. They cannot instantly shift by employing MB understanding. Likewise, individuals do not leave a dominant social media app once it becomes choked with advertisements, even if alternatives exist. Likewise, sports teams, dance partners, and jazz bands all need to practice to accommodate changed circumstances of their performance.

We propose a model in which convention formation is treated simultaneously as joint exploration and collective action. Due to incomplete information (in the sense of~\cite{harsanyi1967games}), the population must coordinate collective action to explore and discover opportunities for mutually profitable conventions. This is accomplished by agents' incrementally ``learning by doing''~\citep{arrow1962economic}. In this view, the social good produced by successful collective action is the establishment of a convention. We are concerned with the general case of \emph{conflictual coordination}~\citep{vanderschraaf1997joint}. Some potential conventions may be more favorable to one party over another. Our model corresponds to situations in which multiple desirable conventions can in principle be established, but are unknown at the start. Furthermore, while each convention is preferred by different individuals, establishing \emph{any} convention is more desirable for everyone than failure to do so is for anyone. In the model we propose, individuals learn to implement policies that sequence elementary actions over time and space. This is in line with other recent work where multi-agent reinforcement learning has been used to develop mechanistic models that extend to the case of spatio-temporal complexity \citep{eccles2019learning, jaques2018social, koster2020silly, lerer2019learning, mckee2020social, perolat2017multi, peysakhovich2019reinforcement, peysakhovich2019robust, shum2019theory, zheng2020ai}\footnote{In contrast, some multi-agent models of convention emergence resonate with the MF account in certain aspects, e.g. myopic decision making~\citep[e.g.][]{Marchant2017LimitedOA, delgado2002emergence, franks2013manipulating, Gavrilets6068, SHOHAM1997139}, but agents knowing the rules to a small abstract gamespace does not pose an exploration problem.}.

The premises on which our model is based are:
\begin{enumerate}
    \item Large world with incomplete information. 
    
    \item There is more than one Nash equilibrium.
    
    \item Individuals have heterogeneous tastes.
    
    \item Attempting to coordinate bears an opportunity cost.%
    
    \item Individuals learn by retrospective caching of long-run values which they use greedily for decision-making. 

\end{enumerate}

From these premises we show that MF convention formation is indeed possible in our formulation. In accord with the literature on critical mass theory~\citep{marwell1993critical}, simulations of our model show that this group achievement is opposed by start-up and free rider problems. This explicit tension results from collective action requiring coordination while more directly self-interested action does not. It is akin to the mixed motives professional athletes, scientists or politicians face when part of a team with a common payoff while also maintaining a personal brand. The model also allows us to test novel hypotheses on convention formation. We show the emergence of conventions is sensitive to the intensity of taste between groups, variation in intensity of taste within groups, visual salience, the ability to sanction and exclude, and the spatial layout of the environment.


\section{Model}

\subsection{Markov games}

We consider multi-agent reinforcement learning in partially observable general-sum Markov games \citep{shapley1953stochastic, Littman94markovgames}. In each game state, agents take actions based on a partial observation of the state space and receive an individual reward. The rules of the game are not assumed given; agents must explore to discover them. Thus it is simultaneously a game of \emph{imperfect} information---each player possesses some private information not known to their adversary (as in card-games like poker)---and \emph{incomplete} information---lacking common knowledge of the rules \citep{harsanyi1967games}. Agents must learn through experience an appropriate behavior policy while interacting with one another.

We formalize this as an $N$-player partially observable Markov game $\mathcal{M}$ defined on a finite set of states $\mathcal{S}$. The observation function $\mathcal{O} : \mathcal{S} \times \{1, \dots , N\} \rightarrow \R^d$, specifies each player's $d$-dimensional view on the state space. In each state, each player $i$ is allowed to take an action from its own set $\mathcal{A}^i$. Following their joint action $(a^1, \dots , a^N) \in \mathcal{A}^1 \times \! \dots \! \times \mathcal{A}^N$, the state changes obeys the stochastic transition function  $\mathcal{T} : \mathcal{S} \times \mathcal{A}^1 \times \! \cdots \! \times \mathcal{A}^N \rightarrow \Delta(\mathcal{S})$, where $\Delta(\mathcal{S})$ denotes the set of discrete probability distributions over $\mathcal{S}$, and each player $i$ receives an individual reward defined as $r^i: \mathcal{S} \times \mathcal{A}^1 \times \dots \times \mathcal{A}^N \rightarrow \R$.

\subsection{Heterogeneous preferences}

The environment contains berries, which can be consumed by individual players. There are $K$ different berry colors. Each agent prefers consuming any berry over not consuming anything, but also prefers to consume one particular berry over all others (default experimental condition). We use the term `taste' to refer to the rank ordering of instantaneous rewards an agent receives from consuming berries of different colors.

Tastes are usually congruent with  \emph{unconditional preferences}---i.e.~motivations reflected by the behavior of agents when they are alone or have no need to take others into account. \cite{bicchieri2006grammar} distinguishes unconditional preferences from \emph{conditional preferences}, which motivate agents to choose certain behaviors conditionally on the expectation that others will act in a particular way\footnote{\cite{bicchieri2006grammar} distinguishes between empirical and normative expectations. The former concerns expectations of what others will do. The latter concerns expectations of what others think should be done. While normative expectations are relevant in the broader study of social norms, only empirical expectations are relevant to convention~\citep{bicchieri2006grammar}.}.

In conflictual coordination games, conventions cause unconditional and conditional preferences to diverge. Concretely, in our model this means that a convention has been achieved if agents pursue a different group objective (i.e. plant a different berry color) than predicted by their taste alone. In the following we will define this more formally.

Let $\mathcal{I}:\mathcal{S} \times \mathcal{A} \times \{1, \dots N\} \rightarrow \{0, 1\}$ be the indicator function that is $1$ when player $i$ consumed a berry in a given state and $0$ otherwise. We consider environments with $K$ different kinds of berries, each represented by a different color, and $K$ corresponding index functions $\mathcal{I}_k$, one for each berry color. We use the notation $\vec{\mathcal{I}}$ to indicate the vector with $k$th entry $\mathcal{I}_k$.

An individual-specific \emph{reward profile} $g$ maps consumption events (as represented in $\vec{\mathcal{I}}$) to instantaneous rewards (determining the individual's `taste'). The members of a population may have different reward profiles from one another. We use the notation $\mathcal{G} = \{g_1, \dots, g_{|\mathcal{G}|}\}$ to indicate the set of unique reward profiles in the population. In principle $|\mathcal{G}|$ could be equal to the population size $N$, but $|\mathcal{G}| < N$ in all cases we consider since some reward profiles are shared by multiple individuals. Whenever there is no ambiguity we drop the subscript and refer simply to a reward profile $g$.

Let $\mathcal{E}$ be a set of Nash equilibria. For an equilibrium $E \in \mathcal{E}$, let $\mu_E$ denote the distribution of states generated from playing out games with each player implementing their part of the equilibrium.

The expected value of equilibrium $E$ to an individual with reward profile $g$ is
\begin{equation}
    U_g(E) = \sum_{s, a} g(\vec{\mathcal{I}}((s, a)) \mu_E(s, a).
\end{equation}

Define the \emph{group objective} $G$ shared by all players with reward profile $g$ as the set of equilibria with expected value to players with reward profile $g$ at least as high as achieved for all other equilibria in $\mathcal{E}$. That is, $G =\text{argmax}_{\mathcal{E}}(U_g) =  \{E~|~U_g(E) \ge U_g(E_k) ~~ \forall E_k \in \mathcal{E}\}$. In this paper we consider conflictual coordination settings, i.e.~where $\bigcap_{g \in \mathcal{G}} G_g = \emptyset$; individuals who do not share the same reward profile prefer different equilibria to one another.

For simplicity, we refer to color-based `groups' of players that gain higher reward for consuming the same colored berry and therefore share a group objective. For example, all players that gain the highest reward for consuming red berries are in the `red group'. Any of these equilibria result in one group profiting more from the achieved consensus than the others, causing tension between the groups.

\subsection{Individual learning behavior}

Each agent learns, independently through its own experience of the environment, a behavior policy $\pi^i : \mathcal{O}^i \rightarrow \Delta(\mathcal{A}^i)$ (written $\pi(a^i|o^i)$) based on its own observation $o^i = \mathcal{O}(s,i)$ and reward $r^i(s,a^1,\dots,a^N)$. Each agent's goal is to maximize a long term $\gamma$-discounted payoff defined as follows:
\begin{equation}
V_{\vec{\pi}}^i(s_0) = \mathbb{E} \left[ \sum \limits_{t=0}^{\infty} \gamma^t r^i(s_t, \vec{a}_t) | \vec{a}_t \sim \vec{\pi}_t, s_{t+1} \sim \mathcal{T}(s_t, \vec{a}_t) \right] \, .
\end{equation}

Each agent has its own private neural network. That is, each neural network is trained using only the unique experience distribution generated by exactly one agent. 

All agents play in all episodes. This protocol has been called independent multi-agent reinforcement learning \citep{lanctot2017unified, laurent2011world, leibo2017multiagent}. See the appendix for details of the simulation, agent architecture and training regime.

\begin{figure}[ht!]
    \centering
    \includegraphics[width=9cm]{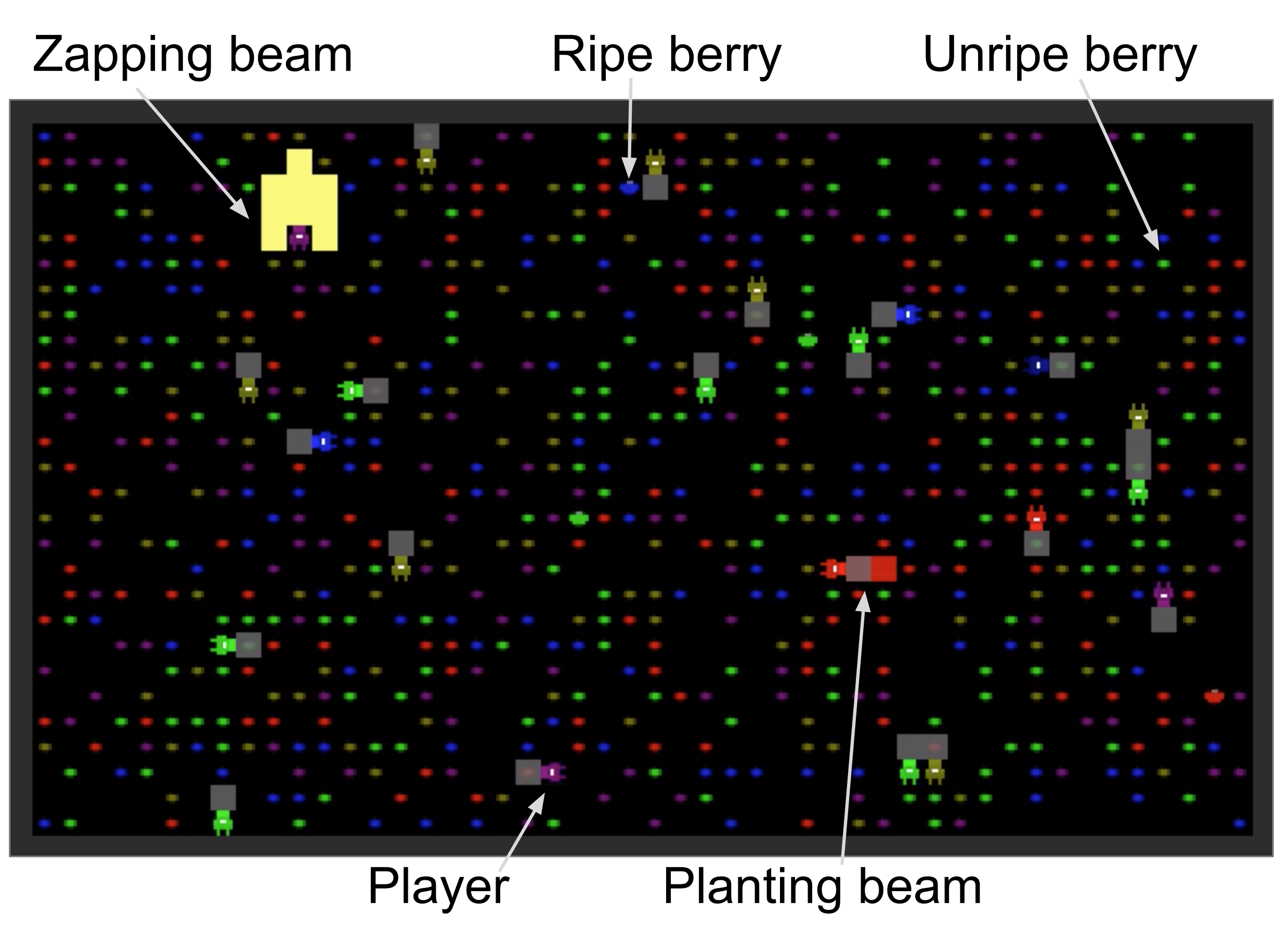}
    \caption{Figure \ref{fig:environment}: Depiction of the environment. The $N=24$ agents inhabit a grid world controlling one player each.  Players earn reward for eating ripe berries (thereby turning them unripe). Each player gets a higher reward for consuming one particular berry color (2 rather than 1). The players can use a planting beam to turn any unripe berry into any color. If they do, the players' avatar turns the same color of the planting beam used. The probability of a berry turning from unripe to ripe grows larger with the number of berries of the same color. The players can use a zapping beam to temporarily remove other players from the game. At the start of every episode the berry colors are evenly distributed, creating an overall low regrowth rate. The environment offers five highly prosperous equilibria where all berries are turned into one color. Based on their reward-profile, groups of player have a specific preference for an equilibrium (in the default setting, players are evenly distributed over taste for 4 of the 5 berry colors). Players can increase their prosperity by achieving a convention, when they coordinate to support one berry color.}
    \label{fig:environment}
    
\end{figure}

\section{Definition of convention}

The extent to which a group's behavior is governed by convention is a matter of degree. The group produces a distribution of joint behavior which we may compare to the distribution of behavior we would expect it to produce in its maximally uncoordinated state. The more different these two distributions are, the more coordinated the group is. The joint behavior upon which they coordinate is called a convention.

However, you cannot choose just any index of behavior to make this comparison. Rather, it must satisfy certain properties (to be explicated below). Given such a \emph{signature} $\sigma: \mathcal{S} \rightarrow \Delta^{K-1}$ we can define the degree of conventionality of a population of $N$ agents with reward profiles $\mathcal{G} = \{g_1, \dots, g_{|\mathcal{G}|}\}$ and joint policy $\vec{\pi}$ as follows. Assume that $|\mathcal{G}|>1$ and $\bigcap_{g \in \mathcal{G}}
G_g = \emptyset$ (conflictual coordination). Let $\mu_{\vec{\pi}, s_0}$ be the distribution of states generated by playing out the population's joint policy $\vec{\pi} = (\pi_1, \dots, \pi_N)$ starting from state $s_0$. The expected empirical signature is the following mixture distribution:
\begin{equation}
    \mathbb{\Sigma} = \sum_{s\in \mathcal{S}}\sigma(s) \mu_{\vec{\pi},s_0}(s).
\end{equation}

The distribution of behavior we expect when the system is maximally uncoordinated can be predicted by considering the \emph{taste-seeking} joint policy $\vec{\pi}_{TS}$. It is the joint policy obtained when each agent pursues its own objective singlemindedly i.e., without regard to opportunity cost (see below). The expected taste-seeking signature is the following mixture distribution:
\begin{equation}
    \mathbb{\Sigma}_{TS} = \sum_{s\in \mathcal{S}}\sigma(s) \mu_{\vec{\pi}_{TS},s_0}(s).
\end{equation}

The degree of conventionality $C$ can then be defined as:
\begin{equation}
    C = \mathcal{M}(\mathbb{\Sigma}_{TS}, \mathbb{\Sigma})
\end{equation}
where $\mathcal{M}$ is any metric for comparing two distributions.

This definition of conventionality relies on two concepts we have not yet adequately explained: (1) how to define the signature function, and (2) how to calculate the taste-seeking signature distribution. Both will always have to depend on features of the environment being modeled\footnote{That the definition of the signature must depend on the details of the environment is not surprising since it plays a similar role in our model as the correlation device plays in other models of convention formation (e.g.~\cite{gintis2010social}). Interestingly though, while correlation devices are seen as entirely exogenous to game theory proper, signatures may be endogenously caused by forces within the multi-agent system itself, potentially rendering them far less mysterious.}.

To be a valid signature, a random variable must satisfy:
\begin{enumerate}
    \item Causally downstream from agent actions.
    \item Insensitive to adding an additional player as long as they elect a free-riding strategy.
    \item For each reward profile $g$, a valid signature maps all $s \in G_g$ to a corner of the simplex. States reflecting different equilibria map to different corners of the simplex.

\end{enumerate}

\begin{sloppypar}
A valid signature for the environment we consider in this paper is the $K$-dimensional vector of proportions of berries in each given color. Let $b_{k}(s) = \frac{\text{number of type \emph{k} berries}(s)}{\text{total number of berries}}$. So ${\sigma(s) = (b_1(s), \dots, b_K(s))}$. It is causally downstream from agent actions since agents in this environment can choose to plant berries of any color. To see that it is insensitive to free riding, note that a sensible definition of a free rider in this environment is an agent that consumes berries, but never plants any itself (see environment details below). Such an agent will have no effect on the berry color distribution. Finally, that the states in $G$ map to corners of the simplex can be seen by noting that these states must be monoculture states, i.e.~states where a single berry color has been planted in every available location. If some berries were not the same color then the agents preferring the dominant color could improve their returns by planting more of it, contradicting that $G$ collects states corresponding to a Nash equilibrium. Therefore $\sigma = (b_1, \dots, b_K)$ is a valid signature.
\end{sloppypar}

We need one additional assumption to predict the expected taste-seeking signature: \emph{equipotency}. It means that all individuals are equally able to effect their objective in principle. Note that this does not mean agents actually know how to achieve their objective at any point in time. Equipotency only requires that nothing would stop them if they worked singlemindedly to cause states in their objective $G$ (without regard for opportunity cost). Let $N_g$ be the number of individuals with reward profile $g$ in the population, so ${N_{g_1} + \cdots + N_{g_{|\mathcal{G}|}} = N}$. Given the equipotency assumption and the signature's insensitivity to free riders, we may predict the expected taste-seeking signature to be $\mathbb{\Sigma}_{TS} = \frac{1}{N} (N_{g_1}, \dots, N_{g_{|\mathcal{G}|}})$. 

In summary, we may assess the emergence of convention in simulation by comparing the empirical distribution over berry types to the distribution we would expect based on the population's intrinsic taste profile, a proxy for the most uncoordinated outcome distribution. For example, a convention emerges when individuals in smaller color groups forego their own taste to work instead toward achieving the objective corresponding to the taste of individuals in a larger color group.

\section{Results}

\texttt{Allelopathic Harvest} is a mixed-motive game played by $N = 24$ players (see Fig.~\ref{fig:environment} and the appendix for details). It combines elements of public goods games \citep{camerer2011behavioral} with elements of conflictual coordination games like Bach or Stravinsky (also called Battle of the Sexes) \citep{luce1957games, osborne2004introduction, snidal1985coordination}. It is embedded in a 2D environment with considerable spatio-temporal complexity where individuals must learn policies to implement their strategic choices.

\subsection*{Game Equilibria}

\textbf{Planting, harvesting and berry growth}

Individuals are rewarded for consuming ripe berries, according to their own reward-profile $g$. There are $K$ different kinds of berries, each with their own color. At the start of each episode, berries are initialized to be unripe and evenly distributed over the $K$ colors. Individuals can replant unripe berries to any color. Thus each individual experiences a tension between their incentive to immediately consume ripe berries and their incentive to plant unripe berries in the color they will prefer to consume upon ripening.

Berry ripening depends (stochastically) on the number of berries sharing the same color that have been planted. Unripe berries of color $k$ ripen with a probability according to $P(\text{ripen}) = F(n_k)$, where $n_k$ is the number of berries of that color. The function $F$ increases monotonically and sharply (it has a positive second derivative), over the interval $[0, 685]$ which covers all possible numbers of berries in the environment (Fig.~\ref{fig:24players}E); see appendix for details. The episodes are very long,  $8000$ frames, (compared to the time it takes to plant and harvest). Thus investing in establishing one color throughout the map ``a monoculture'' is prudent because it can be done relatively rapidly (if all players join in) and by doing so, all players will be able to harvest berries at a much faster rate for the remainder of the episode\footnote{See \textcolor{blue}{\url{https://youtu.be/fw-rAtVD-v8}} for an example episode.}.

Our aim in this section is to characterize the distribution of rewards obtained by agents who roam around their environment harvesting berries. Since the ripening rate of each berry color depends on the number of berries with that color, the expected return of an individual with reward profile $g$ can be decomposed into a sum of terms corresponding to each berry color. The largest of which corresponds to whichever berry color is represented by the most berries. We call the largest fraction of berries sharing the same color in state $s$ the \emph{monoculture fraction} $m(s)$. As $m(s) \rightarrow 1$, the berry ripening rate (for the dominant colored berry type) approaches its maximum value.

\textbf{Characterizing monoculture states}

There are $K$ distinct monoculture states $s^*_1, \dots, s^*_K$ where $m = 1$, one for each berry color. There is a sense in which the joint policies associated with these states are Nash equilibria (see appendix for the precise statement and proof of this claim). Note that there may also be other Nash equilibria besides these. Unlike simple matrix games where actions like cooperate or defect are selected in an atomic fashion, here individuals must implement policies which, when taken as a whole, \emph{amount to} particular strategic choices like cooperation or defection.

Each monoculture state $s^*$ is the group objective $G$ of a group of individuals with reward profile $g$ that ranks the most dominant berry in $s^*$ above all others. Yet the monoculture states are not just tolerated by subgroups with divergent objectives because unilateral deviations are worse. They are also Pareto optimal. Since $F(m(s))$ is optimal when $s = s^*$, total berry consumption is maximized at the monoculture states. Selecting a different monoculture state merely advantages a different group. Selecting a non-monoculture state disadvantages all.

\begin{figure*}[ht!]
		\includegraphics[width=0.8\linewidth]{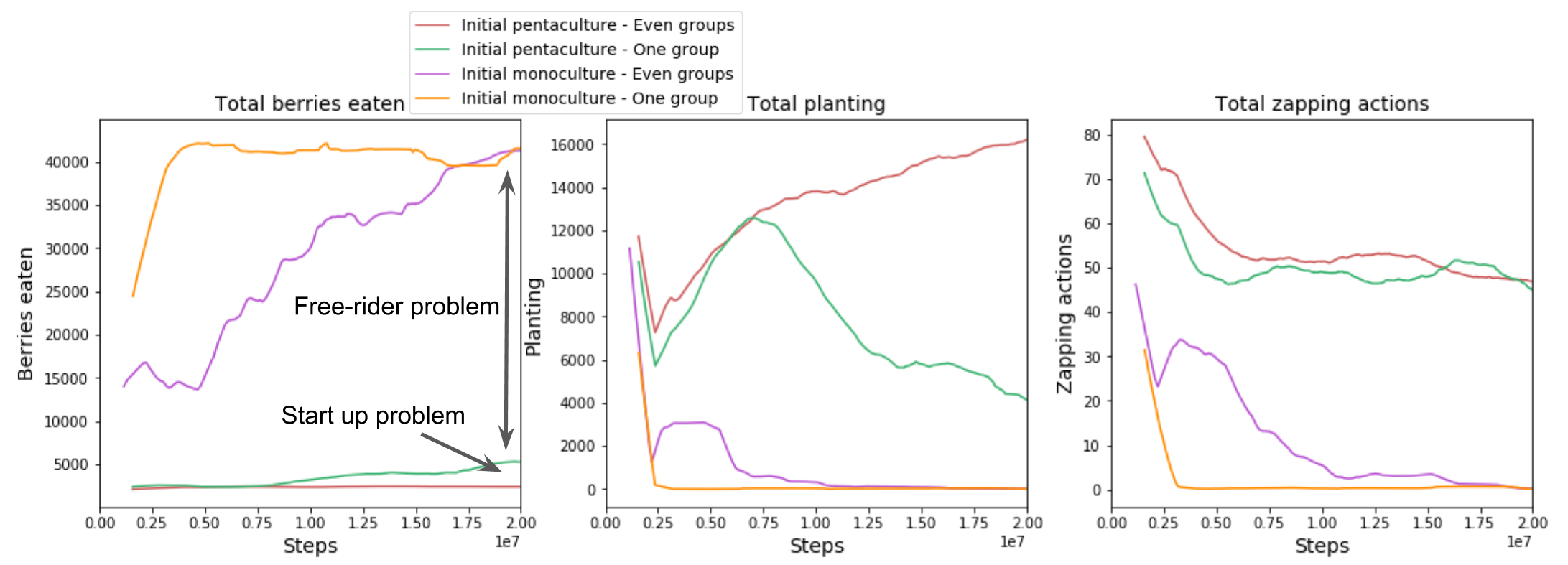}
\vspace{-.1cm}
		\caption{Figure \ref{fig:limits}: The free rider problem, the start-up problem and critical mass. A. Total berries eaten B. Total amount of zapping C. Total planting---all across training in 4 different settings; Even group sizes or all players being in the same group; berries being distributed uniformly at the start of each episode or starting with the same color. The difference between all berries being initialized the same color or different colors reveals the free-rider problem, as the population is unable to achieve the ideal outcome of all berries being the same color. When berries need to be replanted, the difference between having one group or competing groups reveals the start-up problem.}
		\label{fig:limits}
\end{figure*}

We empirically characterize the problems faced by individuals in this model by considering a sequence of experimental conditions (see Fig.~\ref{fig:table})\footnote{Note that the choices which color the bigger or smaller groups were was arbitrary (see Appendix for details on each experiment). Runs are considered for an arbitrarily chosen $1e7$ steps unless otherwise stated.}. In the \emph{'Initial monoculture - One group'} setting, all players are in the yellow group and all berries initialized to be yellow. In this case there is no conflict of interest between individuals. This ideal condition demonstrates the empirical maximum reward that can be achieved in this environment (Fig.~\ref{fig:limits}A). 

In the \emph{'Initial monoculture - Even groups'} setting, there are 4 evenly sized groups of players with different reward profiles, each preferring a different color. This discord between groups is quickly overcome and players learn to not alter the initial monoculture. Despite heterogeneous reward-profiles (tastes), if placed in the perfect yellow equilibrium, the population achieves maximal berry consumption and does no zapping or planting (Fig.~\ref{fig:limits}B\&C). Similarly, even when all players are in the red group the red objective is not established or strongly pursued (result not illustrated) if players are placed in the yellow initial monoculture.

Monoculture states correspond to Nash equilibria and are Pareto optimal (also see supplemental materials). This also shows that group objectives (in the sense defined above) are to achieve a (specific) monoculture state. Like checkmate in expert games of chess\footnote{Experts generally resign before checkmate actually occurs.}, even though monoculture states are rarely reached in practice, these distant targets of the collection action exert a magnetism that organizes all other dynamics in this model\footnote{See \textcolor{blue}{\url{https://youtu.be/iYmLd3_DNOI}} for an episode in this setting. Many thanks to \cite{medina2007unified} for the checkmate analogy.}.

\begin{figure*}[t]
\begin{center}
\begin{tabular}{l l l l l}
 Setting & Berry setting & Groups & Resulting conflict  & Reward\\
 \hline
 \hline
 1 Initial monoculture - One group&  $K = 1, m(s_0) = 1$ &  $|\mathcal{G}| = 1$  & No conflict  & Maximal\\ 
 \hline
 2 Initial monoculture - Even groups &  $K = 1, m(s_0) = 1$ &  $|\mathcal{G}| = 4$ & Unequal outcomes & Maximal\\  
 \hline
 3 Initial pentaculture - One group &  $K = 5, m(s_0) = 1/5$ &  $|\mathcal{G}| = 1$  & Free rider problem & Lower\\  
 \hline
 4 Initial pentaculture - Even groups &  $K = 5, m(s_0) = 1/5$ &  $|\mathcal{G}| = 4$ & Free rider \& Start-up  & Lowest   
\end{tabular}
\end{center}
\captionof{figure}{Figure \ref{fig:table}: Environment equilibria and conflicts. Settings 1-3 obviate particular problems posed by the environment to the population. Setting 4 is the default as it poses both a conflict between groups and between each individual in the population.}

\label{fig:table}
\end{figure*}

Collective action is bookended by twin difficulties known as the \emph{start-up} and \emph{free-rider} problems. In the initial phase of collective action, when there are few cooperators, and little work has yet been done to advance toward a social goal, the motivation to defect is fear that one's efforts will be in vain~\citep{granovetter1978threshold,  heckathorn1996dynamics, marwell1993critical}. This fear is not unfounded. Collective action may fail if too few contribute, or if too many ``back the wrong horse''. Symmetrically, the free-rider problem arises in the late phase of collective action, when most of the population is already engaged. There is a temptation to defect because one could obtain the benefit of the social good without incurring the personal cost of cooperation~\citep{olson1965logic}.
  
\subsection*{The free rider problem}

Consider the case where $K = 5$, all individuals in the population share the same reward profile $g$, and the monoculture fraction of the initial state is $m(s_0) = 1/K = 1/5$ (Setting 3 in Fig.~\ref{fig:table}).

Let $\mu_{\vec{\pi}}$ be the distribution of states generated by playing out the population's joint policy $\vec{\pi} = (\pi_1, \dots, \pi_N)$. The expected monoculture fraction is $\bar{m} = \sum_{s\in \mathcal{S}} m(s) \mu_{\vec{\pi}}(s)$.

Because berries are generally available for all (at least when the monoculture fraction is sufficiently high), this model resembles a public goods problem (as in \cite{olson1965logic})\footnote{Though in this case it is not entirely pure since berry consumption is context-dependently rivalrous.}. Policies that contribute to increasing $\bar{m}$ are only favored if the benefit to an individual of increasing $\bar{m}$ outweighs the cost of the time investment in planting. Empirically, this condition is not satisfied after a certain berry ripening rate has been achieved. Beyond a certain level of $\bar{m}$, which we measure to be $\bar{m}^\dagger = 0.58$, individuals are no longer motivated to contribute to the public good by planting berries concordantly with the emerging monoculture. This is true regardless of whether their own objective $G$ coheres with the emerging monoculture or not. All would benefit from increased monoculture. But all would prefer to spend their own time harvesting and rely on others to put in the effort needed to increase the monoculture fraction beyond $\bar{m}^\dagger$.

Empirically, the comparison between \emph{'Initial monoculture'} and \emph{'Initial pentaculture'} (all 5 berry colors equally distributed at the start of each episode) in the \emph{'One group'} setting reveals a free rider problem (Setting 1 vs 3 in Fig.~\ref{fig:table}). Even when there are no competing groups, the individuals fail to realize the full potential of the environment by coordinating to create a monoculture state. Fig.~\ref{fig:24players}E, shows how much more of the berry regrowth rate could be unlocked.

\subsection*{The start-up problem and critical mass}

In line with other accounts of collective action~\citep{marwell1993critical}, this model contains a start-up problem in all cases where the population contains subgroups of individuals with diverse reward profiles, e.g., $\mathcal{G} = \{g_0, \dots, g_n\}$.

The usual treatment of the start-up problem and critical mass effect begins from a production function, tipping, or threshold model~\citep{granovetter1978threshold, schelling1978micromotives, marwell1993critical}. In contrast, our model highlights a different mechanism which nevertheless produces similar predictions. The idea is that whenever $|\mathcal{G}| > 1$ there is disagreement between subgroups as to which monoculture state would be preferred. Each subgroup may prefer a different objective $G$ from all the others, but all could benefit from agreement. In our model, one way to modulate the severity of the start-up problem is to vary the size of the largest group. Being a member of a large group, the likelihood of one's actions towards the public objective being in vain is reduced, thereby incentivizing action.

The between-group coordination problem is revealed in the \emph{'Initial pentaculture - Even groups'} setting, compared to the \emph{'Initial pentaculture - One group'} setting. Four groups compete to replant berries, leading to low returns (Fig.~\ref{fig:limits}A), more zapping (Fig.~\ref{fig:limits}B) and large amounts of replanting (Fig.~\ref{fig:limits}C). Removing the competition between groups with all players sharing the same reward-profile in the \emph{'Initial pentaculture - One group'} setting leads to higher returns, less zapping and more replanting. When there are competing groups, the population struggles to make strides towards any group objective. 

\begin{figure*}[ht!]
    \centering
    \includegraphics[width=\linewidth]{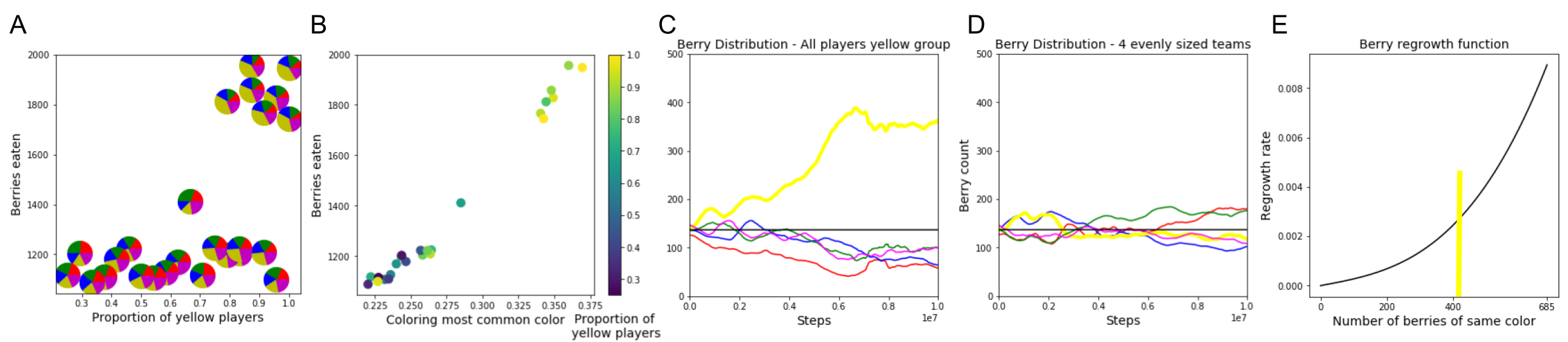}
    \caption{Figure \ref{fig:24players}: A. Berry consumption as function of group size, displaying the resulting berry distribution of each run. B. The relationship of group size and planting the most common color on berries eaten. C./D. Berry distributions over training illustrating runs with all 24 players in the yellow group and 4 groups with 6 players, respectively. E. The function governing the regrowth rate of the berries in dependence of how common the berry color is in the environment, marking the value achieved in C.}
    \label{fig:24players}

\end{figure*}

Additionally, we show in Fig.~\ref{fig:24players} that contribution to replanting is severely limited below a threshold in size of the largest group. Beyond that threshold, contribution levels rapidly transition from vanishingly small up to the level needed to achieve monoculture fraction $\bar{m}^\dagger$ (and after this point the free-rider problem dominates the analysis).

We measure the population's success (measured via berry consumption) as a function of what proportion of the population is in the largest group (yellow, all other players are equally distributed over the red, green and blue group). Each marker in Fig.~\ref{fig:24players}A shows the distribution of berries of each run. Populations can enter two regimes: one in which the pursuit of one equilibrium reaches a critical mass and returns are high, and runs in which no berry color is spread far above the others and returns remain low. Having a high proportion of individuals in the largest group appears to be a necessary but not sufficient condition for entering the high return regime. As can be seen in the color distribution of the markers, the 7 runs with high berry consumption all achieved a relatively wide spread of yellow berries, consistent with most players being in the yellow group. Fig.~\ref{fig:24players}B shows the tight relationship of the populations success and the amount that was contributed towards the public good (planting the currently most common berry color). Fig.~\ref{fig:24players}C and D illustrate the berry distribution over training for runs in which all players are on the yellow group, or evenly distributed. Fig.~\ref{fig:24players}E illustrates the fraction of monoculture achieved in C on the berry regrowth function.

Notice that for the case of $|\mathcal{G}| = 2, K = 2, m(s_0) = 1/K = 1/2$, our model reduces to one that is strategically equivalent to the Bach or Stravinsky game. The difference is that our model requires agents to learn policies to implement their choices rather than simply selecting them in a one-shot manner.

\subsection*{Long term, incremental convention formation}

Based on our empirical results, this model has at least $2K + 1$ stable states: the base conflict state $(\bar{m} \approx 1/K)$ in which no berry is particularly widespread, $K$ incomplete monoculture states $(\bar{m} \approx 0.58)$ where the population has overcome the start-up problem, but remains stymied by free-riding, and $K$ monoculture equilibria states $(\bar{m} \approx 1.0)$, not attainable from most initial conditions we considered. We regard all of these as conventions.

In order to explore the dynamics of convention formation, we consider longer training runs in this section. We find that the system may converge, but also occasionally transitions between stable states representing different conventions. In the runs we consider here $|\mathcal{G}| = 4$, $m(s_0) = 1/K$, and $K = 5$ (the initial state included one berry color that no agents preferred). The four groups consisted of 11, 5, 4, and 4 individuals respectively. 

\begin{figure*}[ht!]
    \centering
    \includegraphics[width=\linewidth]{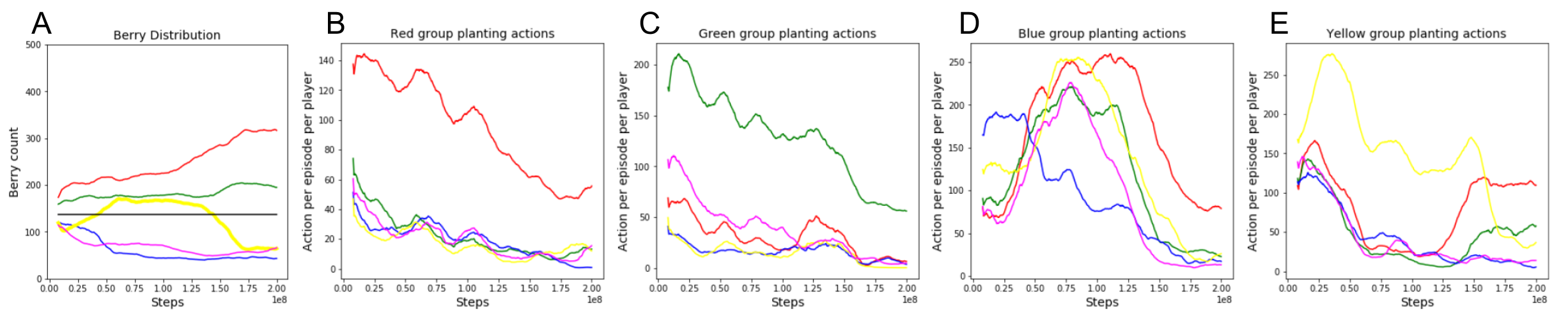}
    \caption{Figure \ref{fig:long}: Conventions developing over long trajectories of training. A. Berry distribution over time. B-E. Replanting behavior of each group. We can observe that while all groups initially support their own color, there is dynamic shifting of the replanting in favour of the red group objective.}
    \label{fig:long}
\end{figure*}

Although the red group contains less than half of the population (11 players), it is able to shift the berry distribution to red over time (Fig.~\ref{fig:long}A). We also observe that first green and yellow are maintained as secondary colors until support for yellow collapses in favor for red. First, each group learns to support their own color (Fig.~\ref{fig:long}B-E). Second, the blue and later the yellow group learn to support red. This demonstrates that individuals sometimes support a convention contrary to the one suggested by their reward-profile, and their doing so encourages others sharing their taste to do the same. Additionally, the overall amount of planting continuously decreases. This suggests that players learn to overwrite each others' actions less (see the supplemental materials for similar results obtained with different seeds).

\section{Additional Results}

\textbf{The effects of heterogeneous taste preference intensity}

\begin{figure}[ht!]
		\includegraphics[width=0.6\linewidth]{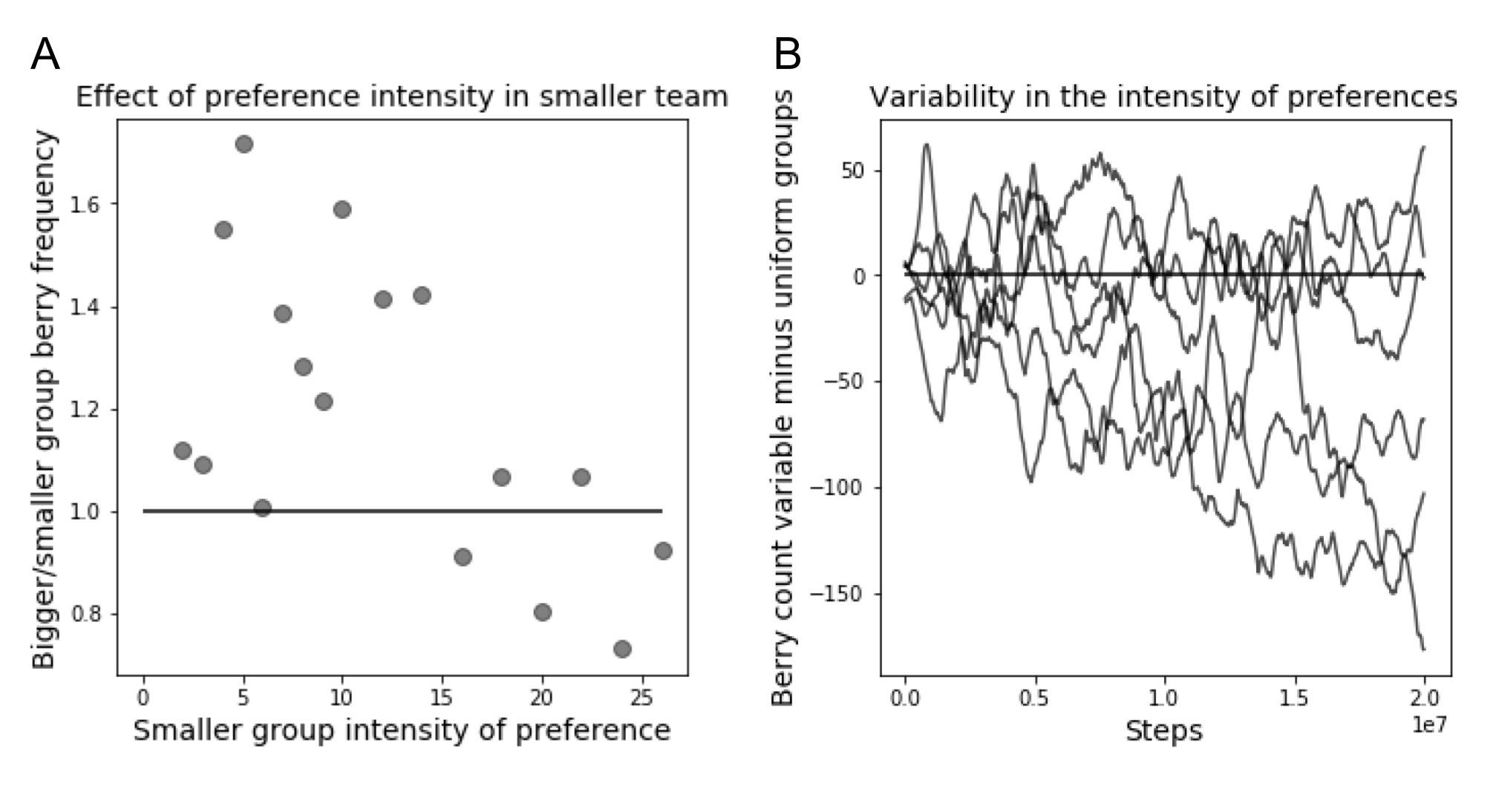}
\vspace{-.5cm}
		\caption{Figure \ref{fig:varied}: A. Berry distribution outcomes over different runs with a larger group with a mild taste preference small group with a strong preference (strength on the x axis). B. Difference in berry amounts over the course of training for colors supported by groups with uniform intensity of taste preference versus groups with varied intensity. Groups with uniform intensity of taste preference across players generally do better.
		\label{fig:varied}
	}
\end{figure}

Taste preferences may not only vary in terms of which group objective is preferred, but also in the intensity of relative reward between that state and others. We consider a case with $|\mathcal{G}| = 2$. The larger group, with objective $G_{g_0}$, consists of 16 players. The remaining $8$ players are in the other group, with objective $G_{g_1}$. Let $g_0(\vec{\mathcal{I}}(s, a)) = 2$ when its most preferred berry is consumed at $s, a$ and $1$ for all other (non-preferred) berries. In each run, we alter the reward that the smaller group gains for each preferred berry. In Fig.~\ref{fig:varied}A, we plot the ratio of berries supported by the two groups, as a function of the intensity of relative reward between preferred and non-preferred consumption by the smaller group. The results show that with very high intensities, the smaller group can shift the berry distribution towards its group objective.

Intensity of taste preference for the same group objective can also vary within a group. Individuals who have a particularly strong interest in achieving a certain collective goal might act as first movers and disproportionately affect the outcome \citep{marwell1993critical, heckathorn1993collective}. We consider a setting with $|\mathcal{G}| = 3$ with $8$ members each. The players in the uniform intensity groups all gain $3$ reward for their preferred berry color. Players in the varied intensity group get between $2$ and $5$ reward for their preferred berry (the mean is the same between groups). In Fig.~\ref{fig:varied}B, we plot the difference between the berry amounts achieved by the varied group and the average of the two uniform groups over learning in different runs. Contrary to our prediction, groups with uniform taste preference intensity across members are more successful at establishing their group objective. This suggests that in order to gain critical mass a group needs to act homogeneously and that the role of specialization in this model is relatively small compared to other multi-agent reinforcement learning-based models of public goods scenarios \citep{mckee2020social}.

\subsection*{Conventions as focal points}

Some group objectives can be of elevated prominence \citep{schelling1960strategy}. When there is a critical mass effect, actions that increase the salience of their allies' contributions can have an outsized influence on the eventual equilibrium selected. For example, news coverage of ``astroturfed'' (i.e., fake) political rallies can tilt the scales in favor of equilibria that would otherwise not have enough support \citep{lyon2004astroturf, walker2014political}.

\begin{figure*}[ht!]
    \centering
    \includegraphics[width=\textwidth]{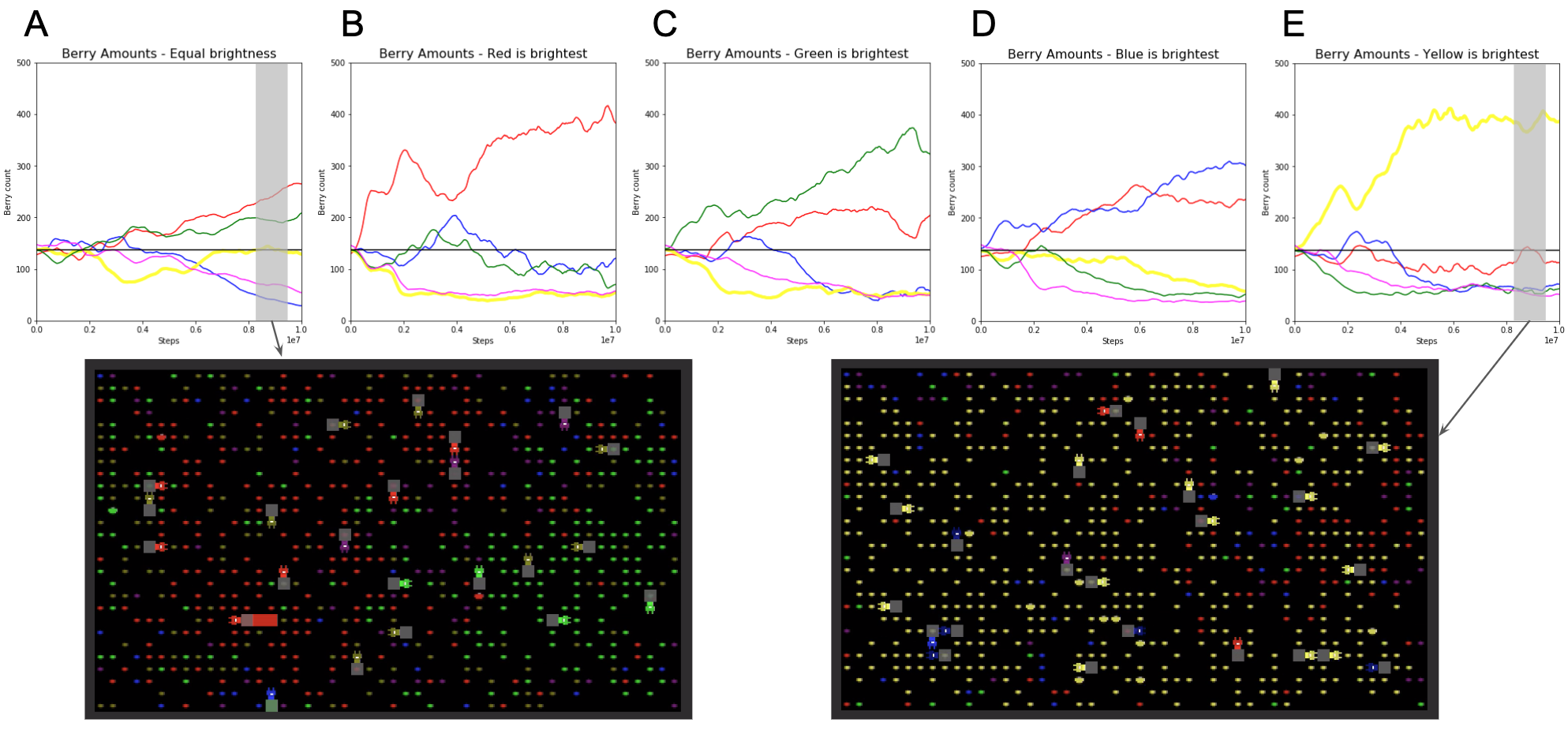}
    \caption{Figure \ref{fig:brightness}: Berry distributions by color over the course of training, dependent on which color is set to be brighter than the others. The black line shows the baseline amounts of berries of each type at the start of every episode. Panel A and E are illustrated with a snapshot of the environment towards the end of an episode late in training.}
    \label{fig:brightness}
\end{figure*}

Our model allows to investigate the effects of visual salience directly. Previous work has demonstrated that agents can show sensitivity to the salience of stimuli, such as size. This follows from the way convolutional neural networks add up the agents' visual inputs (similar effects were observed and discussed in \citep{leibo2018psychlab}). Here, we vary the brightness (sum of RGB values) of one berry type at a time. With $|\mathcal{G}| = 4$ equal groups, it is berry color brightness that determines which berry type is established during the course of learning (Fig.~\ref{fig:brightness}). This demonstrates that stimulus salience can be used to break the symmetry between group objectives.

\subsection*{The effects of excludability and sanctions}

Theories are divided on the question of whether individuals having the ability to sanction or punish one another supports collective action (e.g.,~\citep{macy1993backward, ostrom1998behavioral} versus \citep{obayashi2018self}). Since the model we consider involves complex spatial and temporal dynamics, it offers novel ways to test hypotheses relating to monitoring and sanctioning and how rivalrous or excludable goods are. For example, individuals can learn to exclude others by zapping them if they are close by. Secondly, the spatial layout of the map (i.e. adding walls) can restrict an individual's ability to evade other zaps, and force them to compete for local berries.

In our model, a convention for collective action is established more thoroughly when individuals are unable to zap nearby individuals. Fig.~\ref{fig:supplemental_long}B shows the berry distributions of a run in which the zapping beam was disabled. Consistent with that, averaging over multiple independent runs, Fig.~\ref{fig:supplemental_long}E shows that the entropy over berry types is reduced compared to the default setting.

Fig.~\ref{fig:supplemental_long}C shows the berry distribution of a run in which the map had walls dividing the map into 4 quadrants (accessible from the middle). As shown in Fig.~\ref{fig:supplemental_long}F, when confined into quadrants players zap each other more. To ensure that the results we obtained are not specific to one particular random initialization, Fig.~\ref{fig:supplemental_long} E and F contain additional runs in the default setting (Fig. Fig. \ref{fig:supplemental_long}A) and with zapping disabled (Fig.~\ref{fig:supplemental_long}B).

Fig.~\ref{fig:supplemental_long}D, E and F also include runs of a 'indifferent' players setting. In this setting, players have no heightened reward for any berry over another. Similar to having all players get the higher reward for the same berry color, this removes the group-tension and turns the game into a pure coordination game. Agents are able to significantly reduce the entropy over berries, yielding a higher regrowth rate and reward.

\subsection*{Inability to rapidly switch conventions}

\begin{figure}[ht!]
    \centering
    \includegraphics[width=.7\textwidth]{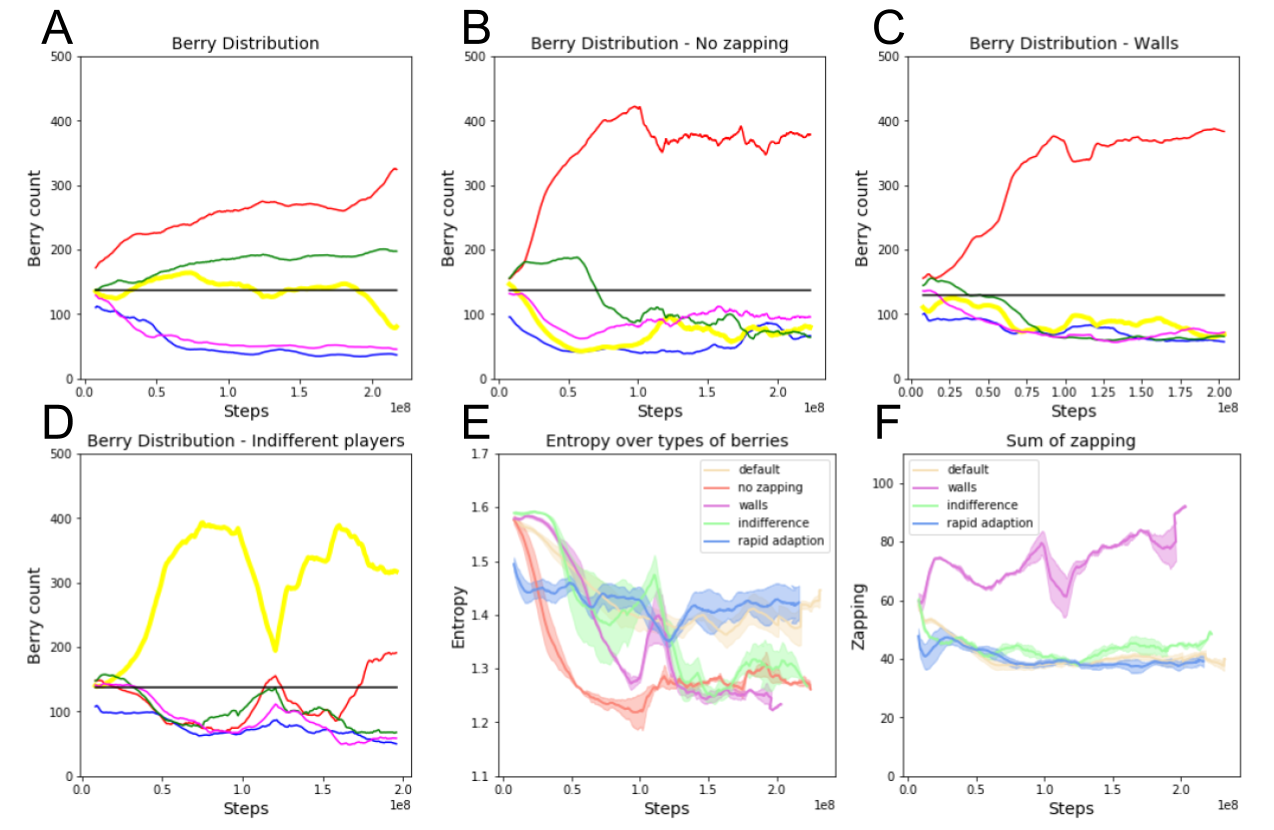}
    \caption{Figure \ref{fig:supplemental_long}: A. Berry distribution over time in default setting with a different random initialization (Fig. \ref{fig:long}). B. Berry distribution over time when zapping is disabled. C. Berry distribution over time with walls in the environment. D. Berry distribution over time with player who get the same reward for all berry types. E \& F shows the average (over multiple independent runs) of the entropy over berry types and sum of zapping across different environment settings.}
    \label{fig:supplemental_long}
\end{figure}

Finally, Fig.~\ref{fig:supplemental_long}, E and F contain results from a setting in which the 24 players did not have unique neural networks. Instead, there was a population of 12 neural networks being trained (5 being rewarded higher for eating a red berry, 3 for blue, 2 for green, 2 for yellow). Each episodes, 24 players were sampled from the 12 neural networks with replacement. This affords the agents an opportunity for meta-learning~\citep{duan2016rl, wang2016learning}: on each episode, the random sampling may favour one particular color (e.g. neural networks that get rewarded more for eating red berries are sampled 16 times on a particular episode). The agents could learn to exploit this random variation and more effectively reduce the entropy of the berry types on the map dynamically, by converging on a different color convention each episode. If the agents fail to learn this, the results are expected to be similar to the default setting. In that case, agents only learn about a convention across episodes (driven the tastes of the underlying population being sampled), but not rapidly adapt each episode---a common behavioral test of MB behavior. The results show that agents are not learning to rapidly adapt their behavior from episode to episode.

\section{Discussion}

Reasoning, habit, and intuition are mechanisms through which an individual may move from thought to thought. Groups of individuals may shift between equilibria as a consequence~\citep{haidt2001emotional}. Theories of convention formation have been largely focused on the effects of reasoning and evolution. Here, we have focused instead on the role of habit as modeled by MF reinforcement learning algorithms. Our model emphasizes the importance of large worlds with incomplete information. Individuals in our simulations begin with no knowledge of their world (their neural networks are initialized randomly). They have to learn from experience everything about their own preferences and affordances in their environment, as well as the rules of the game they play. Joint exploration emerges as the critical mechanism supporting the formation of new conventions. 

It may seem at first that our MF approach shares evolutionary game theory's main limitation: its inability to deal with unique events (see~\cite{gintis2014bounds}). Importantly, unique events do not offer much opportunity for retrospective learning. However, their uniqueness can be mitigated by generalization. MF learning methods like the one presented here, work by caching experienced values after the fact, using general function approximators (deep networks) to represent state-value functions and policies. Agents are capable of operating with novel observations (or states) that were never encountered during training by effective generalization due to learned representations~\citep{mnih2015human, lecun2015deep}.

Our account is focused on the formation of new conventions via MF learning mechanisms. Instead, one might divergently argue that conventions which look habitual were actually built by MB mechanisms. It could be that MB control generally operates in the initial stages of convention formation, but then after a convention is well-established, all individuals adhere to it out of habit (e.g.~\cite{epstein2001learning}): a sort of MB $\rightarrow$ MF consolidation theory. This may well occur for some conventions, e.g., driving on the left or right side of the road\footnote{The MB $\rightarrow$ MF consolidation theory has other potential problems though. For instance, well trained `automatic' behavior (i.e. behavior that is immune to distraction) can still be MB~\citep{economides2015model}.}. However, they are not the kind of convention to which our account applies. Our proposal concerns how MF learning, on its own, may yield conventions. 

We have presented a model that treats convention formation as collective action to explore in large environments of incomplete information and heterogeneous tastes. Our results highlight that convention formation can be further complicated by collective action-related pathologies like the start-up and free rider problems. In particular, we found that populations with large enough pluralities of individuals sharing the same tastes could overcome the start-up problem. But all populations remained stymied by the free-rider problem, and thus were only able to achieve coordination on conventions of intermediate utility. It is possible that overcoming the free rider problem requires higher orders of coordination that MB cognition may afford.

\section{Appendix}

\subsection{Simulation}

\texttt{Allelopathic Harvest} is a mixed-motive game played by $N = 24$ players (see Fig. Fig. \ref{fig:environment}). In the basic setting, the environment (a 50x30 gridworld) is filled with 685 berry plants of 5 colors (reset at the start of each episode; 137 per color). Initially all berries are in an unripe state. Each berry has a probability $p$ to ripen on each step, dependent on the number $b$ of berries of the same color across the whole map; $p = 0.0000025b + 0.000009b^3$ (the 'allelopathic' mechanic (inspired by \citep{leibo2019malthusian}), Fig. \ref{fig:24players} E). These numbers were chosen in order to allow the discovery of the benefits of replanting and entice players to converge on one color to unlock a large payoff.

Players can move around in the environment and interact with berries in several ways. Players can use a planting beam to change the color of unripe berries to one of the 4 other berry-type colors (the 'harvest' mechanic). Players can also walk over ripe berries to consume them. Ripe berries cannot be replanted, and unripe berries cannot be consumed. Players' avatars are recolored after using their planting beam to the same color they turned the berry into. This has the effect that the past planting action remains visible for others until a new action is taken (akin to honest signalling \citep{GINTIS2001103}). We simulated a population of $N = 24$ agents. All agents played in all episodes. This protocol has been called independent multi-agent reinforcement learning \citep{lanctot2017unified, laurent2011world, leibo2017multiagent}. Each player also has a beam that tags out other players ('zapping'). This beam excludes another player for 200 steps from the game if hit, and has a cooldown period of 200 steps. In the set of experiments we describe here, episodes last $T = 8000$ steps. Each agent's visual observation of the environment is limited to a $11 \times 11$ RGB window: 5 units to each left and right, 9 forward facing and 1 backwards. The action space consists of movement, rotation, use of the 5 planting beams, and use of the tagging beam (12 actions total). 

\subsection{Agent architecture and training method}

Each instance of the training regime contained a population of 24 agents, each permanently assigned to a player in the environment. The environment is a gridworld of size $48 \times 28$ units each containing a sprite of $8 \times 8$ pixels. For each timestep $s$, each player $i$ receives an observation of size $88 \times 88$ pixels (i.e., $11\times 11$ sprites) and produces, a policy $\pi^i$, and an estimate of the value $V_{\vec{\pi}}^i(s)$ with a neural network, implemented on a GPU. This neural network was trained by receiving importance-weighted policy updates \citep{espeholt2018impala} sampled from a queue of trajectories. These trajectories were created by 100 simultaneous environments on CPUs that run the environment. The agents received truncated sequences of 100 steps of trajectories in batches of 16. Note, that as each neural network is persistently assigned to the same player, each player maintains its individual experiences and resulting weights throughout training.

For each observation, the neural network architecture computed a policy (probability per action) and value (for each observation). It consisted of a visual encoder, which projected to a 2-layer fully connected network, followed by an LSTM, which in turn projected via another linear map to the policy and value outputs. The visual encoder was a 2D-convolutional neural net with two layers: with 16, and then 32 channels, kernel size 8, and then 4, and stride size 8, and then 1. Notice that the stride size was set equal to the sprite size for efficiency. The fully connected network had 64 RELU neurons in each layer. The LSTM had 128 units. 

We used a discount-factor of 0.99, the weight of entropy regularisation of the policy logits was 0.003. We used the RMS-prop optimizer (learning rate=0.0004, epsilon=1e-5, momentum=0.0, decay=0.99). The agent also minimized a contrastive predictive coding loss \citep{oord2018representation} in the manner of an auxiliary objective \citep{jaderberg2016reinforcement}, which in this case, promoted  discrimination between nearby timepoints via LSTM state representations.

\subsection*{Group composition and colors used}

For the experiments in the main text we consider up to 4 groups that have a taste preference for 4 of the 5 berries, leaving one berry to be a control or baseline (unless there is only one team). All colors were a permutation of the values $[200, 100, 50]$ to control for the mean and variance of RGB values. Note however here that not all colors-values are orthogonal to each other. The same settings were used in Fig.~\ref{fig:varied}A in which there are only 2 groups (16 red players and 8 green players). Note that in all of these experiments, the experimental question is independent of the color values. We consider multiple runs in which we vary the team size or taste preference intensity. In no case is the experimental manipulation confounded with the RGB values.

In Fig.~\ref{fig:varied}B we consider 6 runs, containing each 1 group with varied intensity of taste preference and 2 groups where all players have the same intensity of taste preference (3 for the preferred berry, 1 for all others). Each group contained 8 players. To balance the effect of colors, we picked 3 orthogonal values for the 3 teams ($[200, 0, 0]$, $[0, 200, 0]$, $[0, 0, 200]$) while the berry types that were not preferred by any group had grey values. 

In Fig.~\ref{fig:brightness} the effect of the sum of RGB values of each color is the experimental manipulation. We picked values for red, green, blue, magenta and yellow that add up to 200 in all cases. Four groups were equally sized with 6 members each. When a color is changed to be particularly bright, the sum of RBG values adds up to 450, e.g. $[100, 100, 250]$. As the experimental question concerns the sum of the RGB values, we do not control for variance or correlation between colors in this experiment.

\subsection{Monoculture states are associated with Nash equilibria and are Pareto optimal}

We want to show that monoculture states are Nash equilibria and Pareto optimal. That is, for a suitable notion of cooperation versus defection, we show that unilateral deviations from joint policies that maintain a monoculture state are not profitable (Nash equilibrium). Furthermore, when the population contains heterogeneous taste preferences, the joint policies that maintain monoculture states are Pareto optimal in the sense that no other outcomes can make any individuals better off without also making other individuals (with different taste preferences) worse off.

Let $\Pi$ denote the set of all individual policies. Consider a partition of $\Pi$ into a subset $\Pi^C$ consisting of policies labeled `cooperate', and a subset $\Pi^D$ consisting of policies labeled labeled `defect'. We define this partition by thresholding a scalar behavioral metric function $\alpha: \Pi \rightarrow \R$. See \cite{leibo2017multiagent} for background on this approach\footnote{Unlike simple matrix games where actions like cooperate or defect are selected in an atomic fashion, in the model described here, agents must implement policies which, when taken as a whole, \emph{amount to} particular strategic choices like cooperation or defection. Following \cite{leibo2017multiagent}, we can associate to any policy a scalar index function $\alpha(\pi) \mapsto a \in \R$. Think of $\alpha$ as picking out some feature of the behavior encoded in $\pi$. For instance, \cite{leibo2017multiagent} classified policies by their level of aggressiveness, as measured by how often they used a zapping action. Then by choosing a threshold $\eta$ it is possible to classify policies into those for which $f(\pi) < \eta$ and those for which $f(\pi)\ge \eta$. With knowledge of the expected returns achieved by combinations of policies classified on both sides of $\eta$, you can answer questions like what are the Nash equilibria of the empirical game.}.

\begin{sloppypar}
Let $N$ be the number of individuals in the population. Let ${\vec{\pi}:\mathcal{O}^N \rightarrow \Delta(\mathcal{A})^N}$ denote the joint policy of all $N$ individuals. Multi-agent reinforcement learning is typically developed in the language of joint policies whereas game theory is usually framed in terms of a different concept: strategy profiles. However, there is a deep correspondence between the two notions. We will show that monoculture states are Nash equilibria as defined in terms of unilateral deviations from joint policies rather than strategy profiles, but the argument could easily be translated to the more game theoretic language.
\end{sloppypar}

Individuals are rewarded for consuming ripe berries, according to their own reward-profile $g$ (see main text). There are $K$ different kinds of berries, each with their own color. Berry ripening depends (stochastically) on the number of berries sharing the same color that have been planted. Unripe berries of color $k$ ripen with a probability $P(\text{ripen}) = F(n_k)$, where $n_k$ is the number of berries of that color. We assume the function $F$ has positive first and second derivative (monotonically increasing and concave up).

We define the monoculture fraction $m(s)$ associated to a state $s$ to be the largest fraction of berries sharing the same color in state $s$. As $m(s) \rightarrow 1$, the berry ripening rate (for the dominant colored berry type) approaches its maximum value. Note that there are $K$ distinct monoculture states $s^*_1, \dots, s^*_K$ where $m = 1$, one for each berry color. 

Each monoculture ``state'' is in fact really a set of associated states for which all berries are the same color, but all could be either ripe or unripe, and individual agents could be in any location. Since the monoculture fraction controls everything important about a state in our analysis, we treat each monoculture set as if it were a single state. This is further warranted because all intersections of monoculture states (sets) are empty $(s^*_i \cap s^*_j = \emptyset ~~ \forall i, j)$. That is, whenever $m(s) = 1$, the fraction of berries in all other colors besides the dominant color is zero.

Let $\mu_{\vec{\pi}, s_0}$ be the distribution of states generated by playing out the population's joint policy $\vec{\pi} = (\pi_1, \dots, \pi_N)$ starting from state $s_0$. The expected monoculture fraction for starting from state $s_0$ and playing joint policy $\vec{\pi}$ is $\bar{m}(\vec{\pi}, s_0) = \sum_{s\in \mathcal{S}} m(s) \mu_{\vec{\pi}, s_0}(s)$. In the following we will usually suppress the dependence of $\bar{m}$ on the initial state $s_0$. This is warranted because we always assume that $s_0$ is in the focal set of states $s^*_k$. Thus we study unilateral deviations from joint policies that, having started in a monoculture state, maintain its monoculture status.

When a joint policy $\vec{\pi}$ maintains $\bar{m}(\vec{\pi}) = 1$, we say that it is an `all cooperate' joint policy and its individual policy elements are all in $\Pi^C$. Such a joint policy  maintains the average monoculture fraction at its maximum level. Now, hold all agents but one constant. If one agent now unilaterally deviates, changing from a policy in $\Pi^C$ to a policy in $\Pi^D$. Then, by definition it will cause $\bar{m}$ to decrease\footnote{As per the method of \cite{leibo2017multiagent}, Policies are classified according to a social behavior metric into those that decrease the monoculture fraction from its maximum level $\pi\in\Pi^D$ and those that do not $\pi\in\Pi^C$.}. We make the further assumption that a unilateral deviation can only decrease the average monoculture fraction by an amount $\bar{m}_{\delta} \ll 1$.

Let $k$ be the dominant berry color corresponding to monoculture state $s^*_k$. We assume that agents prefer non-dominant berries by a factor of at most $t$ for which,

\begin{equation}
    t < \frac{F(1) - F(1 - \bar{m}_{\delta})}{F(\bar{m}_\delta)}.
\end{equation}\\

\textbf{Theorem 1: }Given the assumptions above, the monoculture states are Nash equilibria.\\

\textbf{Proof:}\\

Let $\pi^C \in \Pi^C$ denote a cooperation policy and $\pi^D \in \Pi^D$ a defection policy. It is sufficient to compare the joint policy $\vec{\pi} = (\pi^C_1, \dots, \pi^C_N)$ (where all agents cooperate) to unilateral deviations $(\pi^D_1, \pi^C_2, \dots, \pi^C_N)$ (where all but one agent cooperate). Let $g$ be the reward profile of the deviating agent.\\

First consider the case where the deviating agent prefers the dominant berry color. In this case it is clear that any amount of replanting will only lower expected return (the agent spends time coloring and also lowers the overall growth rate). But what if the agent prefers a non-dominant color?\\

There are two possibilities corresponding to replanting in their own preferred color versus any other non-preferred color. We need only consider the case of replanting in their preferred color since the payoff achieved by preferred color replanting is always greater than that of non-preferred color replanting. Let $t$ be the proportional increase in reward per berry consumed of the deviating agent's preferred color versus the dominant color. Any agent implementing such a policy would thus expect their return to increase by $F(\bar{m}_{\delta})t$. However, this increase would happen simultaneously with a corresponding decrease in the rewards obtained from berries of the dominant color. It would diminish from $F(1))$ to $F(1 - \bar{m}_{\delta})$. Since we assumed that $F$ is monotonically increasing and has positive second derivative, we can conclude that $F(1) - F(1 - \bar{m}_{\delta}) > F(\bar{m}_{\delta})$. That is, the missed dominant colored berry growth caused by diminishing the monoculture fraction by $\bar{m}_{\delta}$ is greater than $F(\bar{m}_{\delta})$, the berries produced in the preferred color. Thus, without accounting for $t$, the ``raw'' value destroyed by diminishing the fraction of the most dominant berry by $\bar{m}_{\delta}$ would always be greater than the gains from consuming the agent's most preferred berry color.\\

\begin{figure}
    \centering
    \includegraphics[width=6cm]{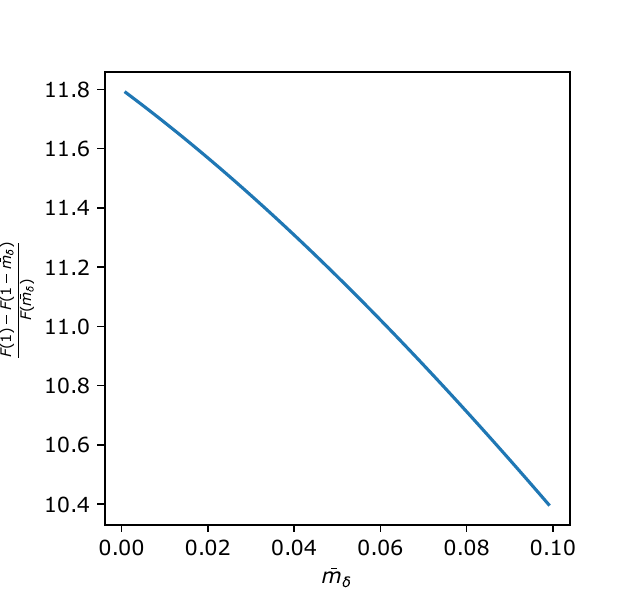}
    \caption{Figure \ref{fig:taste_vs_indiv_monofrac_reduction}: The x axis shows the amount by which an individual can unilaterally reduce the monoculture fraction. The y axis shows the maximum ratio of preferred versus dominant berry reward for which the monoculture state is a Nash equilibrium. When individuals' ability to reduce the monoculture fraction by planting their preferred color is small, the relative taste preference for the individuals preferred color has to be very large.}
    \label{fig:taste_vs_indiv_monofrac_reduction}
\end{figure}

When we take into account the relatively greater reward $t$ obtained per berry when they are of the preferred versus non-preferred color, we find that these berries must be quite significantly more rewarding to harvest\footnote{In our model, unless stated otherwise, we generally use a relatively modest ratio of rewards for preferred versus non-preferred berry colors ($t=2$ for most of our experiments).} to be worth going against the established monoculture. That is, unless $t$ is large enough to compensate the loss of reward from the dominant berry, policies that deviate by reducing the monoculture fraction will not be favored (see Fig.~\ref{fig:taste_vs_indiv_monofrac_reduction}). Thus the monoculture states, $s^*_1, \dots, s^*_K$, are all associated with a joint policy in Nash equilibrium.\\

\textbf{Theorem 2: }Given the assumptions above, and assume further that there are at least two different group objectives in the population, then the monoculture states are Pareto optimal.\\ 

\textbf{Proof:}\\ 

The proof follows immediately from the definition of the group objective $G$ in the main text.\\ 

It suffices to consider monoculture states since increasing the monoculture fraction is always a Pareto improvement.\\ 

Consider a monoculture state $s^*$, when the joint policy is $\vec{\pi}$. The expected value for a group of individuals sharing group objective $G$ is $U_g(s^*) = \sum_{s, a} g(\vec{\mathcal{I}}((s, a)) \mu_{\vec{\pi}}(s, a)$. There are two cases. Either $U_g(s^*)$ is maximal for individuals with group objective $G$ or it is not. If it is not, then there must exist another monoculture state $s^{*\prime}$ (corresponding to a different berry color) for which $U_g(s^{*\prime})$ is maximal. Transitioning from $s^*$ to $s^{*\prime}$ is a Pareto improvement if all individuals in the population share the objective $G$. However, when some individuals do not share the objective $G$ (because they have a different reward profile), then the transition from $s$ to $s^{*\prime}$ is not a Pareto improvement. This holds for all pairs of monoculture states. Thus whenever there are at least two group objectives in the population all monoculture states are Pareto optimal.

\bibliography{biblio}

\end{document}